\documentstyle[aps,prl,multicol,epsf]{revtex}

\begin{document}

\title{Adaptive Random Walks on the  Class of Web Graphs}

\author{Bosiljka Tadi\'c$^*$}

\address{Jo\v{z}ef Stefan Institute,
P.O. Box 3000, 1001 Ljubljana, Slovenia }


\maketitle
\begin{abstract}
We study random walk with adaptive  move strategies on a class of
directed graphs with variable wiring diagram. The graphs are grown from
the evolution rules compatible with the dynamics of the world-wide Web
[Tadi\'c, Physica A {\bf 293}, 273 (2001)], and are characterized by a
pair of power-law distributions of out- and in-degree for each value
of the parameter $\beta$, which measures the degree of rewiring in the graph.
The walker adapts its move strategy according to locally available information
both on out-degree of the visited node and in-degree of target node.
A standard random walk, on the other hand, uses the out-degree only.
We compute the distribution of connected subgraphs visited by an ensemble of
walkers, the average access time and survival probability of the walks.
We discuss these properties of the walk dynamics relative to the changes
in the global graph structure  when the control parameter $\beta $ is varied.
For $\beta \geq 3$, corresponding to the world-wide Web,
 the access time of the walk to a given level of hierarchy on the
graph is much shorter compared to the standard random walk on the same graph.
By reducing the amount of rewiring towards rigidity limit $\beta \to \beta_c
\lesssim 0.1$, corresponding to the range of naturally occurring biochemical
networks,
the survival probability of adaptive and standard random walk
become increasingly similar. The adaptive random walk can be used as an
efficient message-passing algorithm on this class of graphs for large
degree of rewiring.
\end{abstract}

\begin{multicols}{2}

\section{Introduction}
It has been recognized recently that the  key  role in
 the  dynamics of complex systems is played by the
evolving networks of various structures  \cite{Strogatz,Reka}.
Evolution of complex networks representing catalytic reactions,
 cell metabolic processes,  relaxation
in disordered systems,  protein folding, and realistic
 ecological, social, and information
networks \cite{Strogatz,Reka,metabolic,folding,ecology,Mark,SCN,WWW}
are guided by variety of principles inherent to each particular network.
Technically, the  study of the evolving networks includes
theory of graphs, as opposed to regular lattices, which are appropriate
mathematical objects in the study of driven  cellular automata with
diffusive dynamics \cite{DD}. Both, the
dynamics of {\it nodes} and {\it arcs} (links) contribute to the evolution
of the network. Thus, network growth implies that a complex
architecture of links emerges, which  is peculiar to the network, as an
effect of the individual microscopic rules at work.

Many complex networks self-organize  into a scale-free structures of
links (for a recent review see \cite{Reka}).
The appearance of a scale-free
structure is not accidental, having that networks grow from stochastic
rules, that assumes certain degree of disorder on the microscopic scale.
This rules result in the occurrence of dynamic fractal structures
 on the growth time-scale \cite{BT1}, which is  then manifested in the
emergent hierarchical organization of node degrees at large evolution times.
In growing networks the {\it preference} attachment such as in socially
motivated linking, e.g., in science citation network \cite{SCN}, was
shown  to be one of dominant mechanisms \cite{other_mech} leading to the
scale-free structure  of links \cite{Reka,BJA}.
However, in many other growing networks the preference attachment
alone has no capacity to take into account the intimate relationship
 between structural complexity of the networks and their evolution.
This implies dynamically changing  wiring diagram,  in sharp contrast to
the rigid link structure in the science citation network, for example.
Here we concentrate on this type of networks with  the  world-wide Web
as a prominent example, in which constantly updated links yield rapidly
changing wiring diagram  \cite{Strogatz,BT1}.

On the other hand, the occurrence of various {\it universality
classes}---characterized by the same scaling exponents---suggests that
only certain details of the microscopic dynamics are relevant for the
universal scaling behavior of complex networks, in
analogy to critical behavior in physical systems in equilibrium
\cite{critical}. Although the theoretical background of the
{\it universality} in complex dynamic systems and networks is still
missing, numerical simulations on specific models \cite{BT1} and
master equation calculations \cite{DMS,DM,Rodgers1,Rodgers2}
suggest that the following local properties (alone or in the appropriate
combinations) of the dynamics are relevant for the emergent universal
behavior:
(a) Directed vs symmetric links (looked from a node in the network);
(b) Fixed vs variable wiring diagram and time scale of its variations
compared to the growth time scale;
(c) Type of preference linking; and (d) Presence of constraints in the
microscopic growth rules \cite{constraints}.

Recently we have introduced a model of growing directed graph with
bias updates and biased attachment of links, motivated by the conduct
of agents in the real world-wide Web  \cite{BT1}. It was shown that the
model captures minimum relevant
properties for the dynamics of the Web, leading to a  satisfactory
quantitative comparison of the universal scaling quantities  with the
ones  measured in the real Web \cite{WWW}.
The relevant control parameter of the
model $\beta $, defined as the average ratio of updated and added links
per time step (thus measuring the degree of rewiring in the graph)
can vary in the range $(0,\infty )$.

The dynamic processes such as random walk specifically designed to the
Web graphs reveal important information on network's
relaxation upon triggering and suggest how these networks can be
efficiently explored. In Ref.\ \cite{BT2} we introduced  a random walk
with adaptive walk strategy, that utilizes the locally available information
on the  underlying graph structure.
In previous works \cite{BT1,BT2} we have discussed the structure
and dynamics of the world-wide Web, representing a directed graph in this
class with rapidly changing wiring diagram: Estimated
\cite{BT1} value of  $\beta \sim 3-4$ from the scaling agreement with
the observed behavior \cite{WWW}.

In this work we extend the study to the entire class of graphs generated
by varying the control parameter $\beta $ in the physical range $(0,\infty )$.
In the limit of extreme  rewiring ($\beta \to \infty$) some of the scaling
features of the class disappear. On the other end, for $\beta \to \beta _c(N)
\lesssim 0.1$ the graph undergoes a rigidity transition, where majority
of links in the graph remains fixed in time.
Among the networks between these two limits are the world-wide Web and
networks representing a potential range of naturally occurring metabolic
networks and catalytic reactions in open environment \cite{DNA}.
 Specifically, we study (1) how the global graph structure varies with
the control parameter $\beta $ in the entire range before the rigidity
transition, and
 (2) how these structural properties influence the  relaxation of the
network and the dynamics of the random  walk with variable walk strategies.
In addition, study of the random-walk dynamics reveal how the local
clusters participate in the global behavior of the graph. We briefly discuss
the potential application of the random-walk path  in communication processes
on these class of graphs.

The organization of the paper is as follows: In Sec.\ II we present the
growth rules of the class of Web graphs. We grow large networks of
$5\times 10^6$ nodes to determine the distributions of out- and in-degree
for the range of values of the control parameter $\beta $. In Sec.\ III we
define the adaptive random walk on this class of graphs and study
the structure of connected subgraphs scanned by an ensemble of such walkers,
which we then compare with the statistics of a standard random walk on
the same graphs. In Sec.\ IV we determine the survival probability
distribution of these two types of random  walk and their access time to a
given hierarchy level.
Sec.\ V is devoted to a short summary and conclusions.

\section{Growth and Structure of the Web Graph}

A directed Web graph is grown from the dynamic rules proposed in Ref.\
\cite{BT1}, which are based on the criteria (a)--(c) listed above.
(Potential effects of aging and  other constraints
on the microscopic dynamic rules are left out of  this work).
A growth step is defined by adding a node and subsequently creating and/or
removing links in the entire network,
that results in total increment $M(t)$ links at time step $t=i$. Naturally,
we take $M \equiv {\overline{M(t)}} > 0$, suggesting that, in the average,
the  number of links increase at each time step. The links are updated in the
following way. An outgoing link originates from a node $n < i$ with the
probability

\begin{equation}
p_1(n,i) =     {{\alpha M + q_{out}(n,i)}\over{(1+\alpha)M*i}}  \ ,
\label{out_linking}
\end{equation}
which is determined by the number of outgoing links $q_{out}(n,i)$
previously created at that node.
The link points towards the node $k < i$ with the probability
 \begin{equation}
p_2(k,i) =      {{\alpha M + q_{in}(k,i)}\over{(1+\alpha)M*i}} \ ,
\label{in_linking}
\end{equation}
where $q_{in}(k,i)$  is the current  number of
incoming links collected at target node up to the step $i$.  It is
assumed that at the time of addition of a node $i$ to the network
$q_{out}(i,i)=q_{in}(i,i)= 0$.  In addition a fraction
$f_0(t)\equiv \alpha M(t)$ of new links are outgoing links
from the new added node $i=t$. Whereas the remaining
$f_1(t)\equiv (1-\alpha )M(t)$ links are the updated links at preexisting
nodes in the network. Hence, the relevant parameter in the model
 is the ratio of updated and added links at
each time step, i.e., $\beta \equiv f_1(t)/f_0(t) =(1-\alpha )/\alpha$,
(independent of the actual increment $M(t)$), which measures the degree
of rewiring (or flexibility) of the graph.
The Eqs.\ (\ref{out_linking})-(\ref{in_linking})
are motivated by the conduct of the agents who create links in the real Web:
A new link is is likely to be created by the most active agents in the network,
and, similarly, the most probable target node will be the one that
already attracted many links.  For consistency, we use $M=1$ throughout
this work \cite{comment-M}.

The numerical simulations in Ref.\ \cite{BT1} demonstrated that the
rules in Eqs.\ (\ref{out_linking})-(\ref{in_linking}) are compatible with the
 observed scaling behavior of the real Web, when the single control parameter
$\beta $ is set to a  given value $\beta \approx 3$, corresponding to
$\alpha \approx 0.2$ within estimated error bars. The dynamically
emerging out- and in-degree cumulative distributions are given by the
power-law functions
\begin{equation}
P(q_{out}) \sim q_{out}^{-(\tau _{out} -1)} \ ;~~~~ \
P(q_{in}) \sim q_{in}^{-(\tau _{in}-1)} \ ,
\label{PP}
\end{equation}
in the asymptotic region for large number of links (i.e., large networks).
The distributions are given in Fig.\ 1 for various values of the control
parameter, showing the $\alpha $-dependent scaling exponents
$\tau _{out}$ and $\tau _{in}$.

The asymptotic scaling behavior of the in-degree distribution that is
generated by the shifted-linear preference rule in Eq.\ (\ref{in_linking})
with $\tau _{in} = 2+\alpha $ was proved exactly in Ref.\ \cite{DMS},
that agrees completely with our results in Fig.\ 1.
 Recently, a similar analysis using rate equation approach was done in Ref.\
\cite{Rodgers1} for a directed graph in which in addition the probability
of creating a link $C(n,k)$
from the preexisting node $n$ to node  $k$ depends on the  out-degree of
the node $n$ and in-degree of the target node $k$.   It was demonstrated that
when $C(n,k)$ is given by a product of two shifted-linear functions of the
corresponding degrees, the emergent distributions of in- and out-degree
are power-law type and  statistically correlated \cite{Rodgers1}.
According to the above rules in Eqs.\ (\ref{out_linking})-(\ref{in_linking}),
the probability for link creation $C(n,k)$
in our model is $C(n,k)= p_1(n,i)\times p_2(k,i)$, which is exactly the type
of  function compatible with a double-power law distributions for
out- and in-degrees, in agreement with numerical results in Fig.\ 1.
Emergence of statistical correlations between in- and out-degree
distributions is  discussed in Ref.\ \cite{BT1}.
The model of Ref.\ \cite{Rodgers1} is slightly different from ours, involving
three parameters that can not be reduced to a single control parameter,
making it difficult to extract an exact value of the exponent $\tau _{out}$
in terms of our control parameter $\beta $ (or $\alpha$). The observed
behavior shown in Fig.\ 1 suggests that $\tau _{out}$ increases
approximately linearly with $\alpha $ for $0.1 < \alpha \leq 0.5$ (i.e.,
$9> \beta \geq 1$).
 However, $\tau _{out}$ increases rapidly in the region below
$\beta \lesssim 1/2$,
suggesting that the distribution of out-degree becomes exponential when
the rigidity of the graph increases towards a finite critical value. Note that
in the ideally rigid graph $\beta \equiv 0$ the out-degree distribution is
trivial: each node has exactly $M$ links that remain fixed in time.

Study of the distribution of connected components done in Ref.\ \cite{BT1}
suggests that for $\beta$ larger than some critical value $\beta > \beta _c(N)$
a giant connected component occurs, similar to the real Web \cite{WWW}.
By decreasing the control parameter $\beta $ the network structure
changes gradually, eventually
undergoing the phase transition into a structure without a giant component
\cite{BT1}. The critical value of $\beta $ depends on the network size $N$
and in the cases studied here it is close to  $\beta _c
\approx 1/12$.
Loss of scaling in the out-degree distribution at  $ 1/3>\beta > \beta_c(N)$,
while the in-degree still exhibits scaling behavior,
 demonstrates the power of cooperation in
the network: Although the rules (\ref{out_linking}) and (\ref{in_linking})
formally look alike, the dynamical variations in $q_{out}(n,i)$ are
due to local force, whereas the increase of $q_{in}(k,i)$ with time is
made by a collective effect. Consequently, certain dynamical processes
which are related to the out-degree structure might change their character
before the network becomes rigid.   Next we study  random walk processes
on this class of graphs.

\section{Random Walks with Variable Strategy}

We define two types of the random walks: a standard random walk,
and  a walk that learns its strategies from the information stored
 at a visited node on the graph, as defined below.
First we grow the graph of $N$ nodes
using the rules in Eqs.\ (\ref{out_linking})-(\ref{in_linking}), and then
start a walk from a randomly selected node, say node $n$.
In the simple case, here called a naive random walk (NRW) for the
 reason to be clear soon, the walker moves along one of the $q_{out}(n,N)$
outgoing links of that node, selecting the link with equal probability
\begin{equation}
w(n) = 1/q_{out}(n) \ ,
\label{wNRW}
\end{equation}
where we write $q_{out}(n) \equiv q_{out}(n,N)$, assuming for simplicity
that the network does not grow  during the walk time. In this way the
 walker selects a target node, say $k$, and moves there making one time
step of the walk.
Note that this standard random walk on the graph of the structure given in
Eq.\ (\ref{PP}) is more complex compared to the case of hierarchical graphs
with a constant branching ratio, in that it moves in an environment with
variable structure of the relevant out-degree.
An adaptive random walk (ARW), on the other hand, makes move selections
with the statistical weight, which is correlated with the linking strategies
of the visited node. In this case we assume that a link $\ell $ from
node $n$ to node $k$ has a weight defined by

\begin{equation}
w_\ell(n)  \sim  p_2(k,N) \ , ~~~\Sigma _{\ell =1}^{q_{out}(n)}
w_\ell (n) =1 \ ,
\label{pw}
\end{equation}
where  $p_2(k,i=N)$ is given in Eq.\ (\ref{in_linking}) with the normalization
indicated in Eq.\ (\ref{pw}). Hence, the adaptive random walker uses the same
type of strategy that was  used earlier by the visited node, thus  moving
preferably along ``hot'' links of the visited node. Note that the
probability $w_\ell (n)$ is not identical to $p_2(k,i)$ both because it is
evaluated at a later time and it is normalized. In this way, the adaptive
walker exploits full information about local graph structure, which is
done by the out-degree of the visited node and in-degrees of all with that node
connected target nodes \cite{comment-q}. The naive random walker,
on the contrary, uses only the first part of information. As a
quantitative measure of different move strategies it is useful to consider
the first-step statistics  \cite{VP-privat}.
We show in Fig.\ 2 the comparative statistics of the {\it first jump} of
an ensemble of $N_a$ walkers with the naive and adaptive strategies,
respectively, simulated
on the same graph and using the same set of random numbers.
It is shown in Fig.\ 2 that an adaptive search of the direction by
the first jump is guided by the graph structure: the slope of the
distribution of the first-jump distances (measured in node ranks)
resembles the in-degree distribution.
For the ensemble of this size the out-degree distances made in the
first jump are correlated with the in-degree distances in view of Eq.\
(\ref{pw}).  In the random search the distribution is still a power-law,
however, the graphs structure is
much less recognized (slope of the distribution of the first jumps
is by one lower than the in-degree distribution of the graphs). Also
much weaker, if any, correlations between in- and out-degree distances
are observed.
 The walker continues
as long as it is visiting nodes with nonzero number of out-links $q_{out} >0$.

\subsection{Structure of connected subgraphs}

A non-interrupted walk scans  a set of {\it connected} nodes on the graph.
The set of visited nodes is thus a connected subgraph smaller than
a connected component that is usually searched
by the Web crawl \cite{WWW,BT1}. An ensemble of random walkers thus can be
used to scan the {\it local structure} of connected subgraphs, which in
principle differs from the global graph structure.

The ``distance'' traversed by the walker on the graph is best measured
in the node-rank differences, i.e., distance in the levels of hierarchy,
in which the graph structure is  nontrivial.
Spatial distribution of nodes is, in principle, random, whereas the
temporal fractal structure  \cite{BT1} occurs on the growing time scale.
In Fig.\ 3 we show the time-integrated probability distributions of
the node-rank distances for adaptive random walk on the graph with $N=10^4$
nodes and varying the control parameter $\beta $. An ensemble of
$N=20\times 10^4$ walkers was employed. In Fig.\ 4 the corresponding
distributions obtained by the ensemble of naive random walkers are shown
that are simulated in the same conditions as the adaptive random walkers
in Fig.\ 3.

 The distributions of connected subgraphs found both by the ensemble of
adaptive, $W(\Delta q)$, and the ensemble of naive, $R(\Delta q)$, random
walkers are power-law type ($\kappa =out,in$):
\begin{equation}
W(\Delta q _{\kappa}) \sim (\Delta q _{\kappa})^{-\delta _{\kappa}} \ ;  ~~~
R(\Delta q _{\kappa}) \sim (\Delta q _{\kappa})^{-\rho _{\kappa}} \ .
\label{Wq}
\end{equation}

Distinction between  the respective distributions for in- and out-degree
distances are resembling the underlying graphs structure. On the other hand,
the difference in the scaling exponents, for instance in $\delta _{out}$ and
$\rho _{out}$ and similarly with  $\delta _{in}$ and
$\rho _{in}$ suggest that ARW and NRW  are two qualitatively distinct
types of processes. Namely, both $\delta _{out} >2 $ and $\delta _{in}>2$,
whereas $\rho _{out}< 2$ and $\rho _{in} < 2$
indicating that the average distance in the case of the naive random walk
diverges \cite{comment-qmax} with $\Delta q \to \infty$ at $N\to \infty$,
whereas the adaptive random walk remains in the confined areas-distances,
$<\Delta q> < \infty$ when  $N\to \infty$. These properties have
important consequences, for instance for the disease spreading \cite{Romu}
{\it along the random walk path} on the graph.

In the case of ARW the similarity between the $W(\Delta q_{out})$ and
$W(\Delta q_{in})$ and the corresponding distributions in Fig.\ 1
in the scaling region is striking.
By increasing the probability $\alpha $ of a link outcoming from the
new-added node, fast
decrease  of the distribution $W(\Delta q_{out})$ in Fig.\ 3  makes the
exponent $\delta _{out}$ difficult to measure for
$\alpha >0.6$ . At a crossover value close to $\beta \sim 1/3$ the
distribution approaches the exponential form.
At $\alpha =1$ all nodes have exactly one out-link, rendering this
distribution trivial.

\section{Survival and Access Time of the Walk}

Next we study the survival probability of a walk and the average access
time of a walker to a given level of hierarchy. These properties
are relevant for
the relaxation of the graph and message passing to a given distance
in hierarchy on the graph, respectively.

The access time is defined as a number of steps necessary for a walker
to make a given distance here measured in the difference of the hierarchy
levels of the departing and accessed node.
In Fig.\ 5 the average access time normalized to the size of the ensemble
$N_a$ is plotted against the difference in the  hierarchy  levels between
departing and visited nodes for two values of the parameter $\beta $ and
distances measured both by in- and out-degree. A remarkable feature is the
power-law behavior of the access time \cite{BT2}. In general, the scaling form
\begin{equation}
<t_{acccess}> \sim (\Delta q_\kappa )^{-\theta _\kappa }f(t/\Delta q_\kappa )
\label{tacc-q}
\end{equation}
applies both for adaptive and naive random walks when $\beta $ is large.
However, the exponents $\theta _\kappa $ ($\kappa$ stands for
``in,A'' and ``out,A'' for the adaptive and similarly ``in,R'' and
``out,R'' for random walk) are both larger than 2 in the case of
the adaptive search   of the destination, whereas  $1<\theta _\kappa <2$
for the naive random walk. Hence, apart from the small distances, the
average  access time of the adaptive random walk is much shorter compared to
the standard random walk.
This makes  the adaptive random walk a
good basis for its potential application as an efficient  algorithm
\cite{search} for message-passing to a given level of hierarchy.
In the case of catalytic reactions, this
property of the adaptive random walk suggests that in the average fewer
number of reactions occur along the path before a targeted reaction at
a given hierarchy level is reached.
Accordingly, the survival probability, $P_s(t)$ shown in Fig.\ 6,
represents two different types of relaxation processes for large graph
flexibility (large $\beta $): a nearly
exponential relaxation in the case of ARW, and stretch-exponential
for the NRW, respectively.

The difference between adaptive and random search for the target
gradually diminishes with increasing rigidity of the graph (i.e., decreasing
the degree of rewiring  $\beta $),
making the wandering on the graphs similar in both cases when the critical
point is approached $\beta \approx \beta_c(N)$. In Fig.\ 6
we show the probability  $P_s(t)$ that  a walk survives for $t$ steps
on the graph, when the control parameter $\beta $ is varied.
At the critical value ($\beta _c\sim 0.081 $ for the graph sizes used in this
simulations), the survival probabilities
of the two types of walks almost coincide statistically.
In the limit $\beta =0$, i.e., when $\alpha = 1$ in Eqs. \
(\ref{out_linking})-(\ref{in_linking}), each node has exactly one out-link,
rendering the
both walk strategies redundant. As we already pointed out, this limit
corresponds to the graph with fixed links, belonging to another
universality class compared to the Web graph that we discuss here.

\section{Discussion and Conclusions}

We have studied a class of directed graphs with variable wiring diagram
caused, for instance, by frequent updates of the out-links in the world-wide
Web while the graph evolves. The relevant control parameter that defines
the class is the average ratio of updated vs. fixed links, $\beta \equiv
(1-\alpha )/\alpha $.  In the case of the Web $\beta $
is estimated \cite{BT1} in the range 3--4, i.e., to each added link in the
Web comes in the average 3 to 4 links that connect preexisting nodes at
each time step of the evolution.
Other members of this class of graphs can be sought in the catalytic
biochemical reactions, in which nodes are substrates (such as ATP,
${\mathrm{CO_2,
H_2O}}$, etc.)  and links are reactions among these substrates
\cite{Strogatz,Reka,BT2}. Since the backward reactions, although
present, are usually driven by much smaller rates than forward reactions
\cite{Bimath_book}, the biochemical reactions are
represented by directed hierarchical graphs, with ATP as a top connected
node. The presence of catalysts or  enzymes selects appropriate reaction
with a high accuracy. Investigations and potential
applications of catalytic reactions among complex molecules is a great
challenge of sciences in the future \cite{DNA}. The control parameter
in this case can be roughly identified as the concentration of the
catalyst, that can vary depending on the type of the reactants and/or
catalyst.

Here we investigated within the growth model of Ref.\ \cite{BT1} how the
structure of the  out- and in-connections of the graph
vary with the control parameter and how that structure influences the
relaxation processes on the graph.
We summarize the properties that appear as direct
consequences of the flexibility of the graph:

(1) In addition to in-degree, the out-degree distribution appears to
be a power-law distribution with a new scaling exponent; Correlations
between out- and in-degree spontaneously develop;

(2) These structural properties of the graphs determine the properties
of the random walks on these graphs. Richness of the doubly-hierarchical
connections allows us to define various types of the random walk
strategies. Specifically, the adaptive random walker fully utilizes
local information on both in- and out-connectivity of a visited node and
adapts its moves to follow the node linking preferences. On the other hand,
a standard random walker makes use of
the out-degree of a visited node searching only the possible ways out.
For large graph flexibility the adaptive strategy proves effective  in the
search of connected subgraphs and in reducing the access time to a given
hierarchy level. Quick relaxation of the graph represented by the ARW path
in this range of parameter indicates  a short list of reactions before a
targeted reaction occur.

The dynamic processes on this class of graphs reflect the underlying
graph structure, leading to a variety of scale-free properties with
distinct functional dependences on the in- and out- node degrees.
Summary of all scaling exponents for varying flexibility (or rigidity)
of the graph is shown in Fig.\ 7.  To understand the theoretical basis of
these dependences an  analytical study in terms of rate equations
of the random-walk  dynamics is necessary.

(3) By increasing the graphs rigidity (reducing degree of rewiring $\beta $)
the advantage of the strategy ``reading full information on the way'' is lost:
Survival of the adaptive  random walk becomes statistically closer
to the one of the naive walk strategy;  Access time  gradually approaches
the Poisson character; At a critical flexibility $\beta _c \lesssim
0.1$ the graph structure undergoes the rigidity percolation transition
to another universality class characterized with predominantly fixed
wiring diagram, in which the random walk dynamics becomes trivial.

(4)The other limit $\beta \to \infty$, representing the extreme
flexibility,  is formally interesting as a possible course of the world-wide
Web evolution.   In this limit the evolving network  {\it at each growth
stage} shares some similarity with the static random graphs! We detect
certain numerical instability in the distributions
that can be related to the changes in the linking probabilities
Eqs.\ (\ref{out_linking})-(\ref{in_linking}) when $\alpha \to 0$.
After a large number $N_1$ nodes are added, these probabilities are
$N_1$-dependent as $p_\kappa \sim \alpha /(1+\alpha )N_1 +
1/(1+\alpha )N_1^{(1-1/\tau_{\kappa})}$, where we assumed that
maximum degree varies with the network size approximately as $q_\kappa \sim
N_1^{1/\tau _\kappa}$, (as before, $\kappa $ stands for ``in'' or ``out'').
 In the random-graph  theory the leading inverse linear term of linking
probability is compatible with the occurrence of cycles (i.e., triangles,
squares, pentagons, etc.) of all sizes, whereas the additional inverse
sublinear term introduces admixture of full graphs \cite{RG_book}. By letting
$\alpha \to 0$ the leading linear term disappears, inducing sudden change
in the structure of subgraphs---now being only full graphs. Now, having
$\tau _{in}\sim 2$ for small $\alpha $ leads to $p_{in} \sim 1/N_1^{1/2}$,
that is compatible with the occurrence of fully connected  pentagons.
Whereas  $\tau_{out}\sim $3--2.5, leading to $p_{out}\sim
1/N_1^{2/3}$--$1/N^{3/5}$ allowing appearance of  at most fully connected
squares. This suggests why the
exponent of the out-degree should stay larger than the one of in-degree
distribution when $\alpha \to 0$.
Numerical simulations suggest (see Fig.\ 7) that $\tau_{out}\to 2.5$,
whereas $\tau_{in}\to 2$ in this limit.

Random walk path may serve as a communication channel in modeling of
packet transport on this class of graphs.  In addition to channels,
the communication processes depend on the properties of packets to be
transported and of driving rate resulting in load at individual nodes.
By varying these parameters on a hierarchical tree it was shown in Ref.\
\cite{Diaz} that a continuous transition to a congested regime occurs,
that exhibits several universal features.
Looked in this context, our study concentrates on the properties of the
communication channels only, and we do not specify any details on
the character of packets. The simulation conditions thus correspond to
idealized low driving rate below the transition point.
Apart from the occurrence of cycles (closed loops), the Web graph
differs from the hierarchical tree in that linked nodes are
{\it preferably} at large distances of hierarchy, in contrast to the
tree structure where only next hierarchy levels are coupled.
We have demonstrated that due to the complex double-hierarchical
structure of the Web graph the channel selection can be done using
various strategies, each of them resulting in a nontrivial statistical
properties of the channels. How these channel properties would influence
the communication processes when more realistic packets and driving
conditions are considered remains for the future study.

In conclusion, we have demonstrated that flexible wiring diagrams in
the class of directed graphs  induces  a number of structural and relaxation
properties that are crucial  both for evolution   of these graphs
and for design of dynamic processes for exploring  their structure.

\acknowledgments
This work  was  supported by the Ministry
of Education, Science and Sports of the Republic of Slovenia. I thank
to Vyatcheslav Priezzhev for helpful suggestions.

\narrowtext
\begin{figure}
\epsfxsize=82mm\epsffile[58 70 508 714]{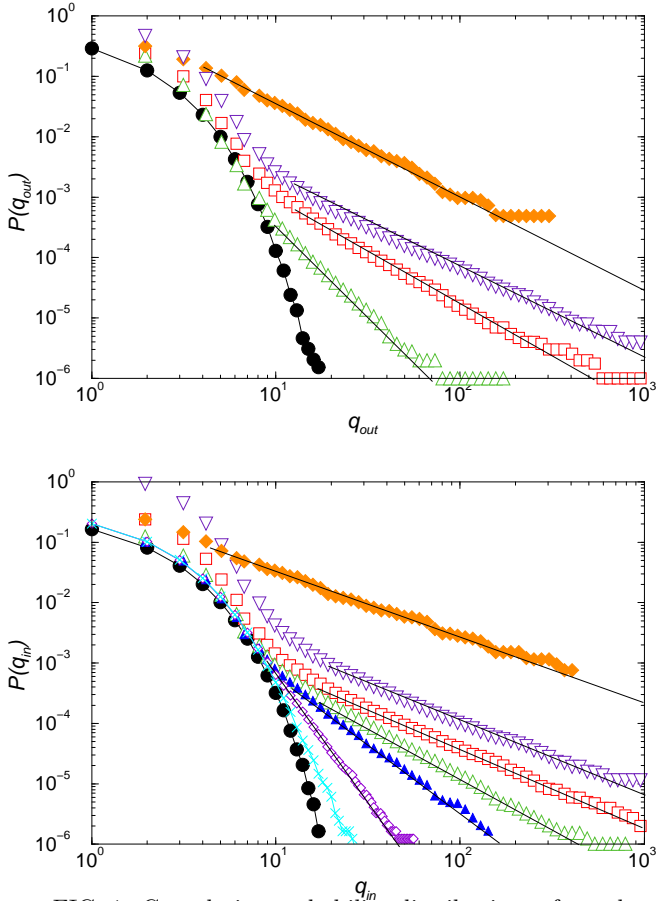}
\caption{\label{fig1}Cumulative probability distributions of out-degree
(top panel) and in-degree (lower panel) of the Web graphs containing
$5\times 10^6$ nodes for varying rigidity of the graph $1/\beta \equiv
\alpha/(1-\alpha )$. Top four curves are for $\alpha =$ 0.0125, 0.125, 0.25,
 and 0.5 (top to bottom in both panels), and  additional three curves
in lower panel for $\alpha =$1, 3, and 12. Line with filled circles:
Exponential distributions corresponding to the case of random selections of
source and target nodes. All data are log-binned with base 1.1 . Top
three curves are moved upwards for better vision. Lines indicate the
scaling region on each curve.
Variation of the slopes with the control parameter are discussed in the text.}
\end{figure}

\begin{figure}
\epsfxsize=82mm\epsffile[53 62 513 374]{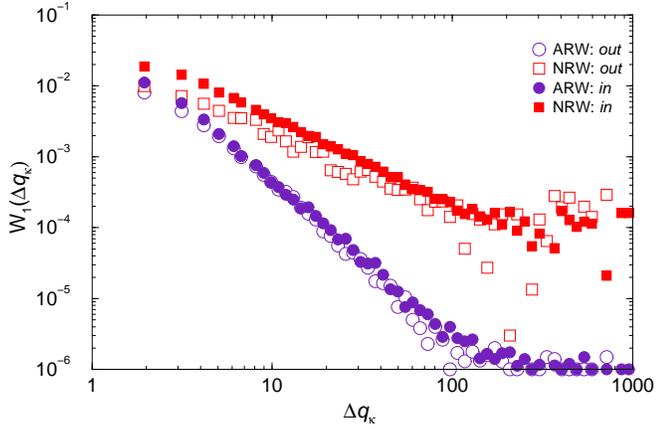}
\caption{\label{fig2} Distribution $F_1(\Delta q_{\kappa})$ of the distance
on hierarchy levels $\Delta q_{\kappa}$ ($\kappa \equiv $out,in) made in the
first jump of the adaptive (ARW) and naive random walk (NRW), as indicated,
obtained by an ensemble of $N_a = 100\times 10^3$ walkers on the graph
of $N=10^4$ nodes. Slopes of in-degree first step distributions are:
2.15 and 1.15 for ARW and NRW, respectively.}
\end{figure}

\narrowtext
\begin{figure}
\epsfxsize=82mm\epsffile[58 70 508 714]{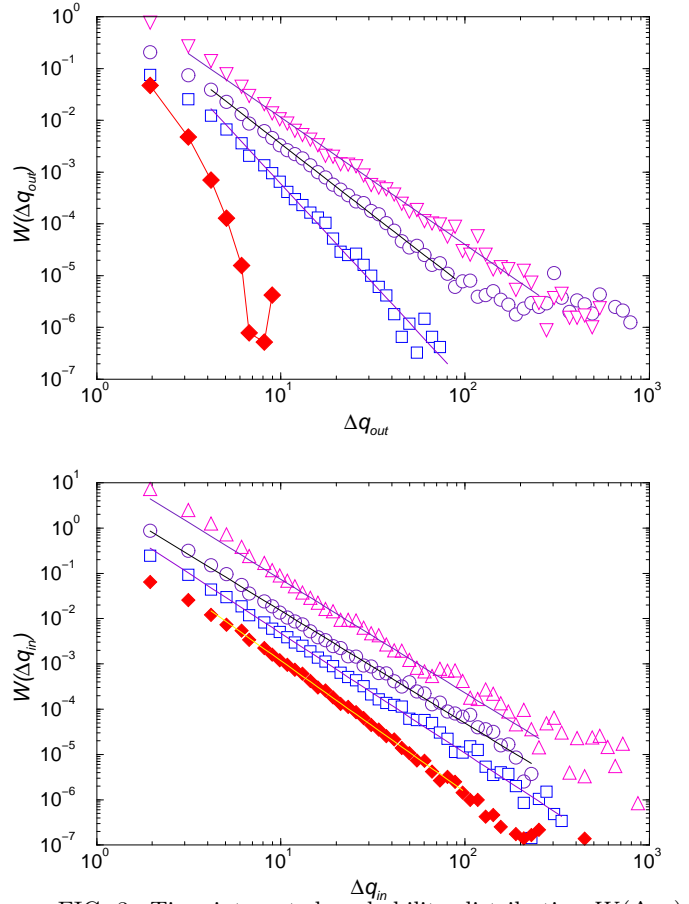}
\caption{\label{fig3}Time-integrated probability distribution
$W(\Delta q_{\kappa})$ of distance in hierarchy levels $\Delta q_{\kappa }$
made by ensemble of $N_a=20\times 10^4$ adaptive random walkers on the Web
graphs of $N=10^4$ nodes for ($\kappa \equiv $out) out-degree distances
(top panel) and ($\kappa \equiv $in) in-degree distances (lower panel).
Different curves correspond to the control parameter values $\beta =$9, 3, 1,
and $\beta _c$= 0.081, top to bottom. Solid lines are pawer-law fits
with the slopes $\delta _{out} =$2.52, 2.75, 3.83, and $\delta _{in}=$ 2.48,
2.50,  2.64, and 2.90, respectively,
 within numerical error bars $\leq \pm 0.06$.}
\end{figure}

\begin{figure}
\epsfxsize=82mm\epsffile[58 70 508 714]{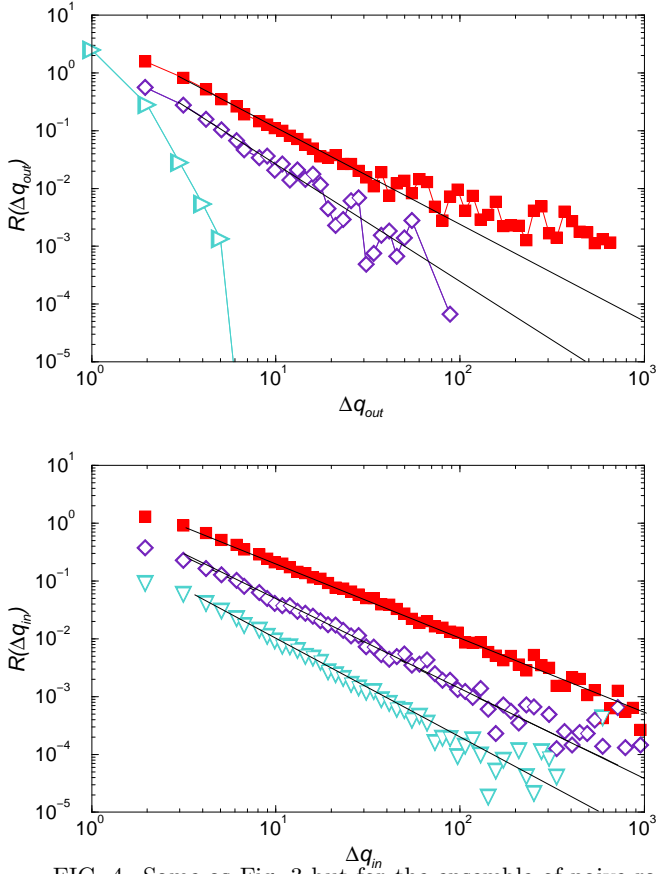}
\caption{\label{fig4}Same as Fig.\ 3 but for the ensemble of naive random
walkers and for $\beta =$3, 1, and 0.081 (top to bottom lines in both panels).
Power-law fits lead to the scaling exponents: $\rho _{out}=$ 1.48 and 2.03,
and $\rho _{in}=$ 1.22, 1.53, and 1.95, respectively,
within numerical error bars $\leq \pm 0.08$.   }
\end{figure}

\begin{figure}
\epsfxsize=82mm\epsffile[46 69 508 336]{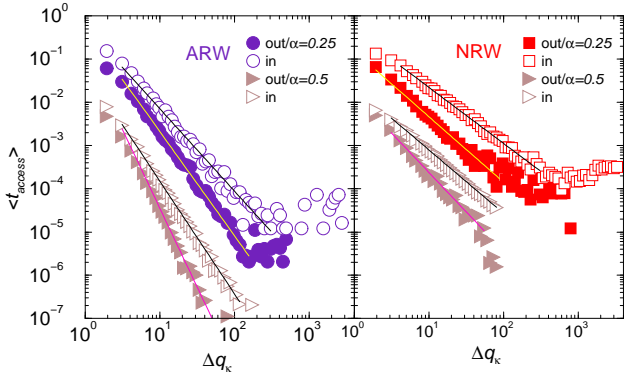}
\caption{\label{fig5} Average access time (normalized by number of walkers
$N_a$) to a hierachy distance $\Delta q_\kappa$ ($\kappa =$ in, out) of
an ensemble of the adaptive (left panel) and naive (right panel) random
walkers for two values of the parameter $\alpha \equiv 1/(\beta +1) $,
as indicated. Distance measured in terms of in-degree (open symbols) and
out-degree (filled symbols). Slopes of the curves top to bottom are: 2.01,
2.41, 2.61, and 3.64 ,  on left panel, and
1.29, 1.51, 1.47, and 1.82, on right panel, respectively.
Estimated error bars within $\leq \pm 0.07$.}
\end{figure}

\begin{figure}
\epsfxsize=82mm\epsffile[45 73 450 336]{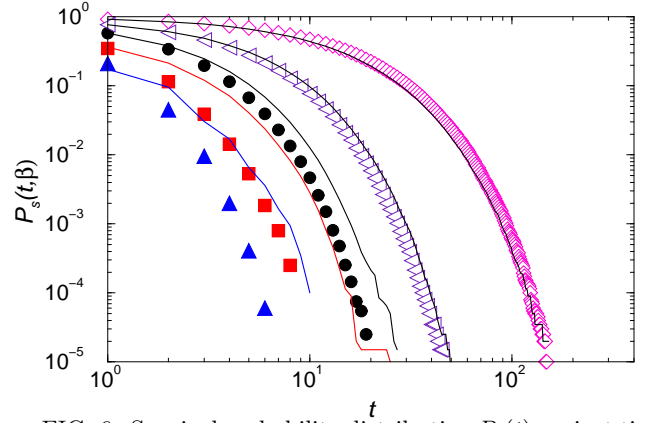}
\caption{\label{fig6}Survival probability distribution $P_s(t)$ against
time steps $t$ for the adaptive random walk (symbols) and respective
naive random walk (solid lines) for varying degree of rewiring $\beta =$ 9,
3, 1, 1/3, and 1/12 (left to right).}
\end{figure}

\begin{figure}
\epsfxsize=82mm\epsffile[70 73 513 336]{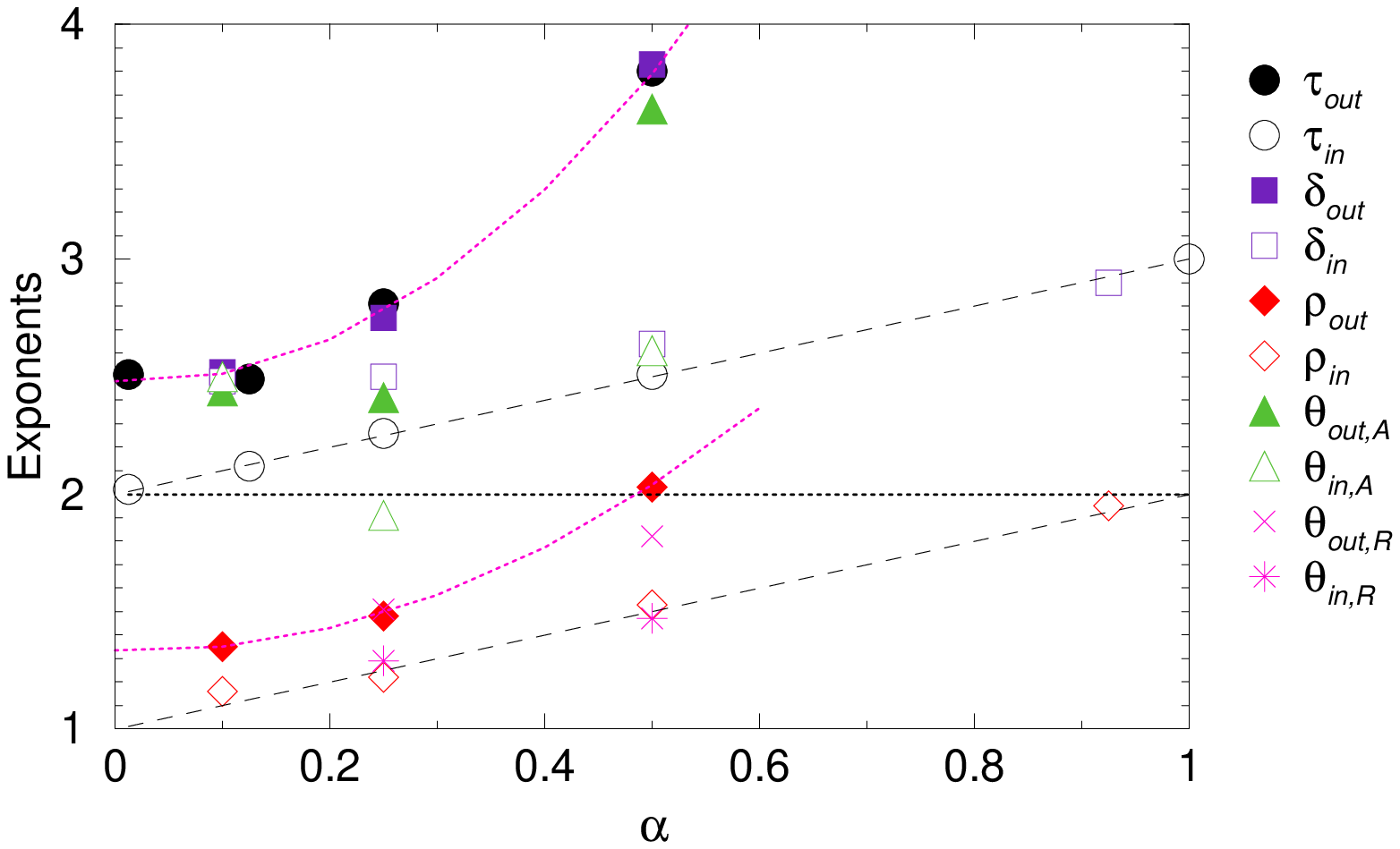}
\caption{\label{fig7}Various scaling exponents ploted vs parameter
$\alpha \equiv 1/(\beta +1)$. Filled symbols correspond to the out-degree
property, and same open symbols---to the corresponding in-degree property,
see legend and definitions in the text. Dashed lines: $2+\alpha $ and
$1+\alpha $. Dotted lines: nonlinear fit of $\tau _{out}$ and
same curve shifted downwards approximately by one. Notice that for the
distributions of the naive random walk we have: $\rho _{in}
\approx \tau _{in}-1$ and $\rho _{out}
\approx \tau _{out}-1$, and for the adaptive walk $\delta _{out} \approx
\tau _{out}$ whereas $\delta _{in}$ deviates from $\tau _{in}$ for small
$\alpha $ values.}
\end{figure}

\end{multicols}

\begin{references}
\bibitem[]{$^*$}$^*$Electronic address: Bosiljka.Tadic@ijs.si

\bibitem{Strogatz} S. H. Strogatz, Nature {\bf 410}, 268 (2001).


\bibitem{Reka} R. Albert and A.-L. Barabasi, ``Statistical Mechanics of
Complex Networks'', Rev. Mod. Phys. (in press).

\bibitem{metabolic} P. M. Gleiss, P. F. Stadler, A. Wagner, and D. A. Fell,
cond-mat/0009124; H. Jeong, B. Tombor, R. Albert, Z. N. Oltvai, and A.
Barabasi, Nature {\bf 407}, 651 (2000).

\bibitem{folding}A. Scala, L.A.N. Amaral, and M. Barth\'el\'emy,
cond-mat/0004380

\bibitem{ecology}J.M. Montoya and R.V. Sol\'e, cond-mat/0011195.


\bibitem{Mark} M. E. J. Newman, Proc. Natl. Acad. Sci. 98,
404 (2001).
\bibitem{SCN} S. Redner, Eur. Phys. J. B {\bf 4}, 131 (1998).

\bibitem{WWW} A. Broder,  R. Kumar,  F. Maghoul,  P. Raghavan,
 R. Sridhar, R. Stata,  A. Tomkins, and J. Wiener, Computer Networks
{\bf 33}, 209 (2000).

\bibitem{DD}D. Dhar, Physica A {\bf 263}, 4 (1999).


\bibitem{BT1}B. Tadi\'c, Physica A {\bf 293}, 273 (2001); cond-mat/0011442.


\bibitem{other_mech} Other mechanisms, for instance in the Internet growth,
may also involve local optimization of supply, see A. Capocci {\it et al.},
cond-mat/0106084.


\bibitem{BJA}A.-L. Barabasi, R. Albert, and H. Jeong, Physica A
{\bf 272}, 173 (1999).

\bibitem{critical} H. E. Stanley, {\it Introduction to Phase Transitions and
Critical Phenomena}, Oxford Univ. Press, (New York) 1971.


\bibitem{DMS} S.N. Dorogovtsev,  J.F.F. Mendes, and   A.N. Samukhin,
Phys. Rev. Lett. {\bf 85}, 4633 (2000).
\bibitem{DM} S.N. Dorogovtsev and  J.F.F. Mendes, Phys. Rev. E {\bf 62},
1824 (2000).


\bibitem{Rodgers1}L.P. Krapivsky, G. J. Rodgers, and S. Redner,
cond-mat/0012181.

\bibitem{Rodgers2}G. Erg\"un and G. J. Rodgers, cond-mat/0103423.


\bibitem{constraints} A special type of aiging, fitness and other constraints
on the microscopic dynamic level are shown to affect the emergent behavior in
the preference attachment model, see discussion in Ref.\ \cite{Reka}.

\bibitem{BT2}B. Tadi\'c, cond-mat/0104029


\bibitem{DNA} Such a reaction in which the reactants are added with
a given rate and the products constantly removed was proposed recently
to study work produced by  molecular DNA motors, B. Yurke {\it et al.},
Nature, {\bf 406}, 605 (2000).





\bibitem{comment-M}The actual number of links in the Web exceeds number
of nodes suggesting that a larger $M$ would be more realistic.
In the model the {\it universal} properties of the network in the scaling
region are not affected when $M$ is varied.


\bibitem{comment-q}Technically, in order to simulate large graphs we
preserve out- and in-degrees at each node from the growth phase and search
links by once again using the rule in Eq.\ (\ref{in_linking}). The potential
differences are minimized with increasing  size of the ensemble.

 \bibitem{VP-privat}V. B. Priezzhev, privat communication.



\bibitem{comment-qmax}The maximum degree scales with
the network size $q_{max} \sim N^D \approx N^{1/\tau }$.



\bibitem{Romu} A different mechanism of desease spreading on hierarchical
graphs was considered in Ref. R. Pastor-Satorras and A. Vespignani,
 cond-mat/0102028

\bibitem{search}It was shown recntly that search time scales
sublinearly with the network size in several other search algorithms that
utilize local information in a power-law graph  with symmetric links, see
 L. A. Adamic, R. M. Lukose,
A. R. Puniyani, and B. A. Huberman, cs.NI/0103016, .






 \bibitem{Bimath_book}J. Keener and J. Sneiyd, {\it Mathematical Physiology},
Springer, Berlin (1998).

\bibitem{RG_book} B. Bollob\'as, {\it Modern Graph Theory}, Springer-Verlag,
New York (1998).


\bibitem{Diaz} A. Arenas, A. D\'{\i}az-Guilera, and R. Guimer\`a,
cond-mat/0009395 .

\end{references}
\end{document}